\documentclass[prb,amsmath,amssymb]{revtex4}

\usepackage{graphicx}
\usepackage{dcolumn}
\usepackage{bm}
\usepackage{color}

\newcommand\Tm{\langle\mathbf{T}\rangle}
\newcommand{\ve}[1][K]{\mathbf{#1}}

\begin{document}

\title{Mean first-passage times of non-Markovian random walkers in confinement}
\author{T. Gu\'erin$^{1}$, N. Levernier$^2$, O. B\'enichou$^2$,  R. Voituriez$^{2,3}$}

 \affiliation{$^{1}$Laboratoire Ondes et Mati\`ere d'Aquitaine, University of Bordeaux, Unit\'e Mixte de Recherche 5798, CNRS, F-33400 Talence, France }
\affiliation{$^{2}$Laboratoire de Physique Th\'eorique de la Mati\`ere Condens\'ee, CNRS/UPMC, 
 4 Place Jussieu, 75005 Paris, France}
\affiliation{$^{3}$Laboratoire Jean Perrin, CNRS/UPMC, 
 4 Place Jussieu, 75005 Paris, France}

\begin{abstract}
\textbf{The 
first-passage time (FPT), defined as the 
time a random walker takes to reach a target point in a confining domain,  is a key quantity in the theory of  stochastic processes\cite{Redner:2001a}. Its importance comes from its crucial role   to quantify the efficiency of processes as varied as  diffusion-limited reactions \cite{benAvraham2000,Benichou:2014},  target search processes \cite{Shlesinger:2006}  or spreading of diseases \cite{Lloyd:2001}. Most  methods to determine the FPT properties in confined domains have been limited to Markovian (memoryless) processes \cite{Condamin2007,Benichou2010,Benichou:2014}. However, as soon as the random walker interacts with its environment, memory effects can not be neglected. Examples of non Markovian dynamics 
 include single-file diffusion in narrow channels\cite{wei2000single} or  the motion of a tracer particle either attached to  a polymeric chain\cite{Panja2010} or   diffusing in simple \cite{Franosch:2011} or complex fluids such as nematics \cite{turiv2013effect}, dense soft colloids \cite{Demery:2014} or  viscoelastic  solutions \cite{ernst2012fractional,mason1995optical}.  Here, we introduce an analytical approach to calculate, in the limit of a large confining volume, the mean FPT of a Gaussian non-Markovian random walker
to a target point. 
The non-Markovian features of the dynamics are encompassed by determining the statistical properties of the trajectory of the random walker in the
future of the first-passage event, which are shown to govern the FPT kinetics.
This  analysis is applicable to a broad range  of  stochastic processes, possibly correlated at long-times.  Our theoretical predictions are confirmed by numerical simulations for several examples of non-Markovian processes including the emblematic case of the Fractional Brownian  Motion in one or higher dimensions. These results show, on the basis of Gaussian processes, the importance of memory effects in first-passage statistics of non-Markovian random walkers in confinement.}
 \end{abstract}


\maketitle

It has long been recognized that the kinetics of reactions is influenced by the properties of the  transport process  that brings reactants  into contact\cite{Redner:2001a,benAvraham2000}. Transport can even be the rate limiting step and in this diffusion controlled regime, the reaction kinetics is quantified by the first encounter properties between molecules \cite{benAvraham2000}. First-passage properties have been studied intensively  in the last  decades  \cite{Redner:2001a,Benichou:2014,Holcman:2014} and are now  well understood when the stochastic motion of the reactants  satisfies the Markov property, \textit{i.e.} is memoryless.    Under this assumption, exact asymptotic formulas characterizing the  first-passage time  of a tracer to a target located inside \cite{Condamin2007,Benichou2010,Benichou:2008a} or at the boundary \cite{Holcman:2014} of a large confining volume have been obtained. These studies reveal that the geometrical parameters, as well as the complex properties of the stochastic transport process (such as subdiffusion), can have a strong impact on the reaction kinetics \cite{Condamin2007,Benichou2010,Benichou:2014}.

However, as a general rule,  the dynamics  of a given reactant results from  its interactions with its environment and  cannot be described as a Markov process.  Indeed, while the evolution of the set of all microscopic degrees of freedom of the system is Markovian, the  dynamics restricted to  the  reactant only    is not. This is typically the case of a tagged monomer, whose non-Markovian motion results from the structural dynamics of the whole  chain to which it is attached  \cite{Panja2010,Guerin2012,Benichou:2015bh}, as observed \textit{e.g.} in proteins \cite{kou2004generalized}. 
Other   experimentally observed examples of non Markovian dynamics  include 
the diffusion of tracers in crowded narrow channels \cite{wei2000single} or in complex fluids such as nematics \cite{turiv2013effect} or  viscoelastic   solutions \cite{ernst2012fractional,mason1995optical}. 
Even in simple fluids, hydrodynamic memory effects and thus  non Markovian dynamics have been recently observed \cite{Franosch:2011}. 
So far,  most of theoretical results  on first-passage properties  of non-Markovian processes have been limited to specific examples\cite{WILEMSKI1974a,Hanggi:1985gf,Masoliver:1986mz,Guerin2012,Benichou:2015bh} or to unconfined systems, where  non trivial  persistence exponents characterizing its long time decay  have been calculated \cite{Krug:1997,Molchan1999,ReviewBray}. However, in many situations, geometric confinement plays a key role in first-passage kinetics\cite{Condamin2007,Benichou2010,Benichou:2014}.  Here, we develop a theoretical framework to determine the mean FPT of non-Markovian random walkers  in confinement. 

More precisely, we consider a  non-Markovian Gaussian stochastic process $x(t)$, defined in unconfined space, which represents the position of a random walker at time $t$, starting from $x_0$ at $t=0$.   As the process is non-Markovian, the FPT statistics in fact depends also on $x(t)$ for $t<0$. For the sake of simplicity, we assume that at $t=0$ the process of constant average $x_0$ is in stationary state (see SI for more general initial conditions),  with increments $x(t+\tau)-x(t)$  independent of $t$. The process $x(t)$ is then   entirely characterized by its Mean Square Displacement (MSD) $\psi(\tau)=\langle[x(t+\tau)-x(t)]^2\rangle$. 
Such a quantity is routinely measured in single particle tracking experiments and  in fact includes all the memory effects in the case of Gaussian processes. At long times, the MSD is assumed to diverge and thus, typically,   the particle does not remain close to its initial position. Last, the process is continuous and non smooth\cite{ReviewBray} ($\langle\dot x(t)^2\rangle=+\infty$), meaning that the trajectory is irregular and  of fractal type, similarly to the standard Brownian motion. Note that the class of random walks  that we consider here covers a broad spectrum of  non-Markovian processes used in physics, and in particular the  examples mentioned above. 

\begin{figure}%
\includegraphics[width=15cm]{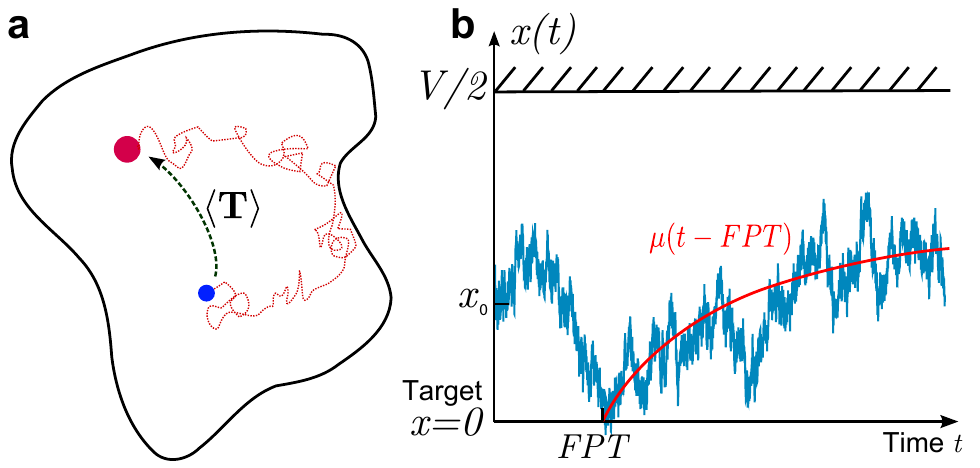}%
\caption{
}
\label{figSketch}%
\end{figure}

The random walker is now confined in a  domain of volume $V$ with reflecting walls, and we  focus on its mean FPT to a  target of position $x=0$ (see Fig. \ref{figSketch}). Note that this setting gives also access to the reaction kinetics of a reactant in the presence of a concentration $c=1/V$ of targets in infinite space. While the theory can be developed in any space dimension (see  SI for an explicit treatment of the 2-dimensional and 3-dimensional cases), it is presented here  for clarity in dimension 1 (see Fig. \ref{figSketch}b).   
Our starting point is  the following  generalization of the renewal equation\cite{Redner:2001a}
\begin{align}
	p(0,t)=\int_0^t d\tau F(\tau) p(0,t\vert \mathrm{FPT}=\tau),\label{RenewalGeneralized}
\end{align}
which results from a partition over the first-passage event. In this equation, $p(0,t)$ stands for the  probability density for being at position $x=0$ at time $t$,  $F$ is the FPT density  and  $p(0,t\vert \mathrm{FPT}=\tau)$ is the probability that $x=0$ at time $t$ given that the first-passage event occurred at time $\tau$. Due to the confinement, $p(0,t)$ reaches for large times the stationary value  $1/V$.  Next, substracting $1/V$ to Eq. (\ref{RenewalGeneralized}) and integrating over $t$ from $0$ to infinity yields an exact expression for the mean FPT  :
\begin{align}
	\frac{\Tm}{V} =\int_0^{\infty}dt [q_{\pi}(t) - p(0,t)], \label{MFPT921}
\end{align}
where $q_{\pi}(t)dx$ is the probability to observe the random walker in the interval $[0,dx]$  at the  time $t$ after the first-passage to the target. 
The exact formula (\ref{MFPT921}) is a generalization of the expression obtained   for Markovian processes \cite{noh2004random,Condamin2007} and holds for any non-Markovian process (even non-Gaussian). Even if $q_{\pi}(t)$ is {\it a priori} a non trivial quantity because it is conditioned by first-passage events, this equation is of great practical use to determine  the mean FPT as we show next.

To proceed further, we (i) consider the large volume limit $V\to\infty$  (where it is assumed that all boundary points are sent to infinity) and (ii) assume that the stochastic process in the future of the FPT, defined by $y(t)\equiv x(t+\mathrm{FPT})$, is  Gaussian with mean $\mu(t)$ and  same covariance as the initial process $x(t)$ (see Fig. \ref{figSketch}b). Simulations and perturbation theory below show the broad validity of this approach.  Eq.(\ref{MFPT921}) then leads to :
\begin{align}
\langle\ve[T]\rangle	=V\int_0^{\infty}dt \ \frac{e^{-\mu(t)^2/2\psi(t)}-e^{-x_0^2/2\psi(t)}}{[2\pi\psi(t)]^{1/2}}.\label{key1}
\end{align}
Relying on a generalization of Eq.(\ref{RenewalGeneralized})  to  link the $n$ times pdfs of $x(t_1), x(t_2), ...$ and the FPT density, we obtain an equation for the probability of the future trajectories $y(t)$   leading to   (see SI for  details):
\begin{align}
	\int_0^{\infty}\frac{dt}{\sqrt{\psi(t)}} \Big\{&\left[\mu(t+\tau)-\mu(t)K(t,\tau)\right] e^{-\mu(t)^2/2\psi(t)}
	-x_0 [1-K(t,\tau)]e^{-x_0^2/2\psi(t)} \Big\}=0 \label{key2},
\end{align}
where $\mu(0)=0$ and $K(t,\tau)=[\psi(t+\tau)+\psi(t)-\psi(\tau)]/[2\psi(t)]$. Eq.(\ref{key2}), which   allows for a self-consistent determination of  the mean future trajectory $\mu(t)$, together with Eq.(\ref{key1}), provide the mean FPT and constitute our main result.

At this stage, several remarks can be done. (i) The mean FPT depends linearly on the confining volume $V$, which extends the  result obtained  for Markovian processes\cite{Condamin2007}. (ii) Our approach reveals the key role played by the mean trajectory $\mu(t)$ followed by the walker in the \textit{future} of the first-passage event. In other words, even if the real motion is stopped at the first encounter with the target, the mean FPT is controlled by the statistical properties of the fictious path that the walker would follow  if allowed to continue   after the first encounter event.  (iii) Assuming that $\psi(t) \propto t^{2H}$ at large times,  with $0<H<1$, it can be shown from the asymptotic analysis of Eq.(\ref{key2}) that 
\begin{align}
	\mu(t)\simeq 	x_0 -A \  t^{2H - 1} \ (t\rightarrow\infty), \label{BehaviorMuInfinity}
\end{align}   
where $A$ is a coefficient depending on the entire MSD function $\psi(t)$ (at all time scales) and on $x_0$ (it generally has the same sign as $x_0$).
Thus, for processes that are subdiffusive at long times ($H<1/2$), $\mu(t)$ comes back to the initial position $x_0$ of the walker, which is consequently not forgotten.  On the contrary,  asymptotically superdiffusive walkers ($H>1/2$) keep going away from the target in the future of the FPT with a non trivial exponent. These behaviors reflect the anticorrelation and correlation of successive steps of subdiffusive and superdiffusive walks, respectively. Note that even for asymptotically diffusive processes ($H=1/2$), $\mu(t)$ tends to a non vanishing constant, in contrast to a pure (Markovian) Brownian motion.  (iv) The importance of non-Markovian effects can  be appreciated by comparing the mean FPT to the result obtained by setting $\mu(t)=0$, which amounts to neglecting the memory of the trajectory before the first-passage.   As shown by Eq.(\ref{BehaviorMuInfinity}),  $\mu(t)$ is actually not small, so that memory effects are significant. These  are especially marked for $H<1/3$, where setting $\mu(t)=0$ in Eq.(\ref{key1}) leads to an infinite mean FPT, as opposed to our finite non-Markovian prediction.

\begin{figure}%
\includegraphics[width=17cm]{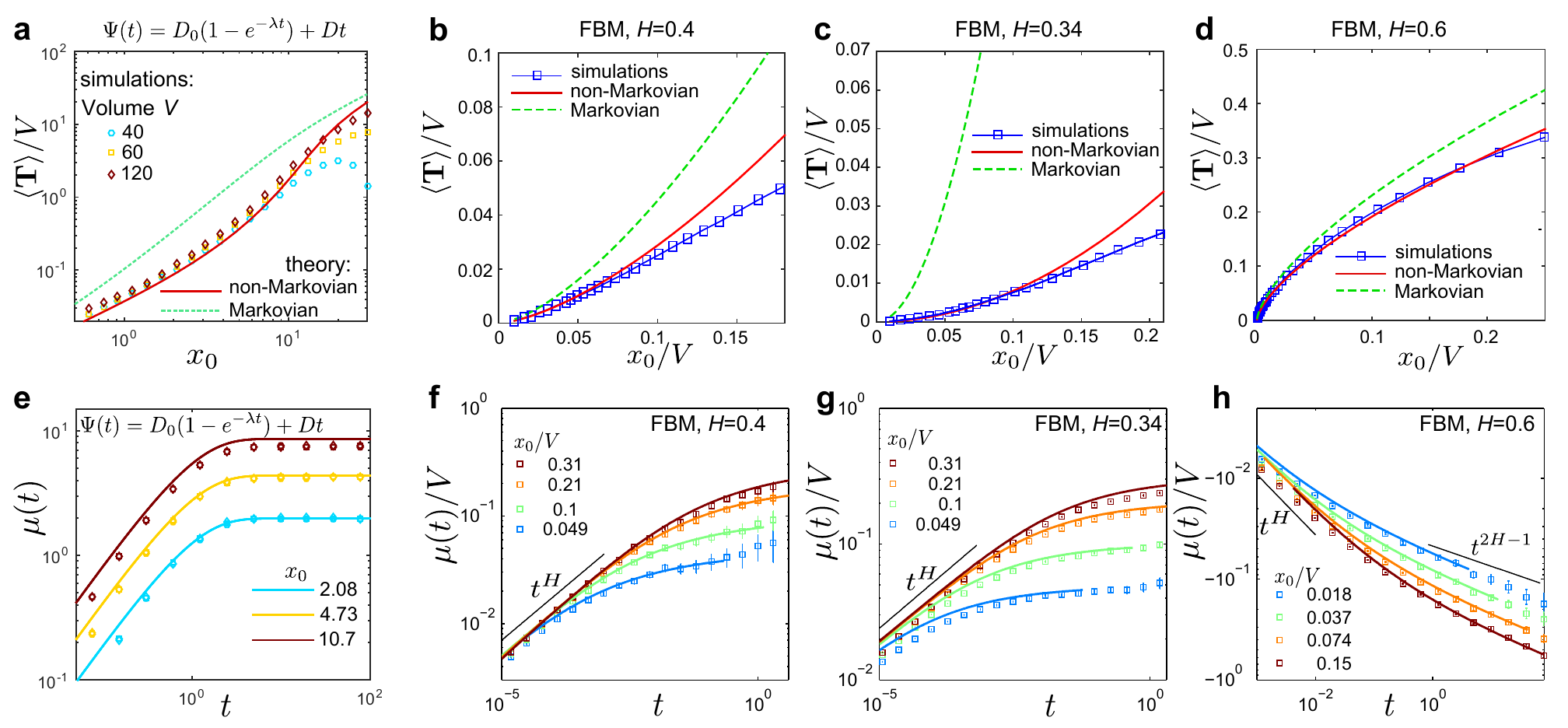}%
\caption{ }
\label{fig1D}%
\end{figure}

\begin{figure}%
\includegraphics[width=14cm]{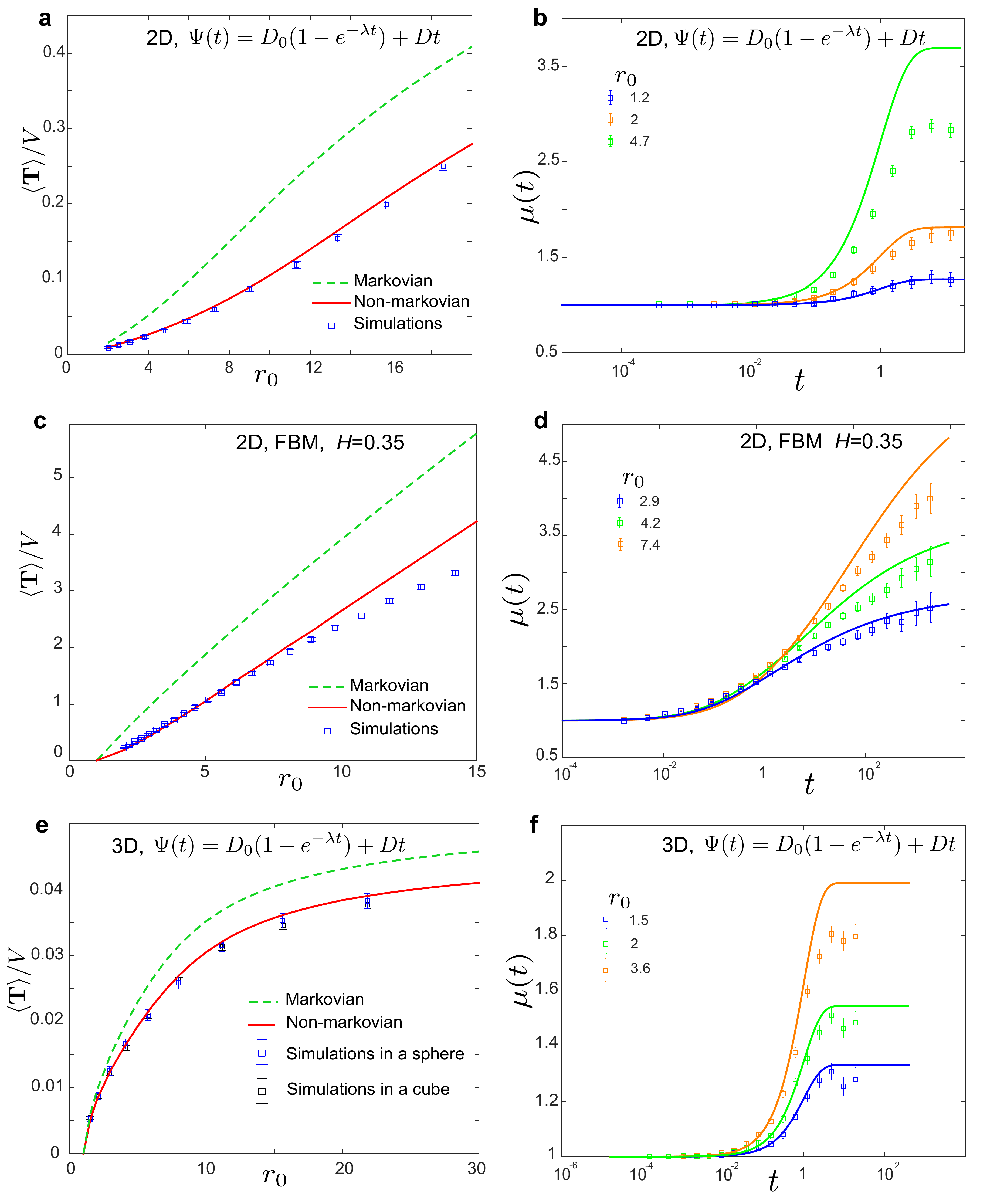}
\caption{}
\label{fig3}
\end{figure}

We now confirm the validity of these analytical results by comparing them to numerical simulations of representative examples of non-Markovian processes defined by the MSD $\psi(t)$. (1) The choice 
	$\psi(t)=D_0 (1-e^{-\lambda t})+ D  t  $
 corresponds to the generic case where the position $x(t)$ is coupled to  other degrees of freedom at the single time scale $1/\lambda$ (Fig. \ref{fig1D}a,e). It is typically relevant to tracers moving in nematics  \cite{turiv2013effect} or solutions of non-adsorbing polymers \cite{ochab2011scale}. (2) The choice $\psi(t)= K t^{2H}$
 where $0<H<1$ and $K$ is a positive transport coefficient (Fig. \ref{fig1D}b,c,d,f,g,h), corresponds to the emblematic Fractional Brownian Motion (FBM) used in fields as varied as hydrology  \cite{mandelbrot1968noah}, finance \cite{cutland1995stock} and biophysics\cite{burnecki2012universal,ernst2012fractional}; it is in particular a good description of anomalous diffusion in various physical situations such as telomere motion \cite{burnecki2012universal} or tracer diffusion in viscoelastic fluids \cite{ernst2012fractional}. This  process is strongly non-Markovian, as shown by its long ranged correlation functions.  For FBM, the solution of Eq.(\ref{key2}) is of the form $\displaystyle\mu(t)=x_0 \ \tilde{\mu}_H\left(t\  K^{1/2H}/x_0^{1/H}\right)$, so that 
 the mean FPT reads
 \begin{equation}
\langle \ve[T]\rangle=V \ \beta_H\  x_0^{1/H-1}K^{-1/2H},
\end{equation}
 with $\beta_H$  a numerical coefficient given in SI. This equation gives the explicit dependence of  the mean FPT on $x_0$ and generalizes the results obtained for Markovian processes \cite{Condamin2007}. (3) The theory can be extended to higher dimensions with the supplementary assumption that the random walk is isotropic ;  two-dimensional and three-dimensional versions of the choices of $\psi(t)$ considered in points (1) and (2) have been analyzed explicitly (Fig. \ref{fig3}). 

Actually, as shown in SI, the theory is   exact at order $\epsilon^2$ when one considers a MSD function of the type $\psi(t)=Dt+\epsilon\psi_1+\epsilon^2\psi_2 +...$ where the small parameter $\epsilon$ measures the deviation to a Markovian process (see SI). Fig. \ref{fig1D},\ref{fig3}  reveal very good quantitative agreement between the analytical predictions and the numerical simulations far beyond this perturbative regime.   
 Both the volume and the source-target distance dependence of the mean FPT are unambiguously captured by the theoretical analysis, at all the length-scales involved in the problem. Note that, even if the theoretical prediction relies on large volume asymptotics, numerical simulations show that it is accurate even for small confining systems (here on the example of cubic shapes).
We emphasize that the very different nature of these examples (space dimension 1,2 and 3,  diffusive, superdiffusive or subdiffusive at long times...) demonstrates  the wide range of applicability of our approach. 
Remarkably,  the amplitude of memory effects is significant in  the examples shown in Fig. \ref{fig1D},\ref{fig3}, where the multiplicative factor between Markovian and non-Markovian estimates of the mean FPT can be up to 15 (Fig. \ref{fig1D}c). As discussed above, this factor is even infinite for the FBM as soon as $H<1/3$. Interestingly, even for
 the process (1) which is diffusive both at short and long times, for which one could thus expect memory effects to be negligible, this factor is not small (typically 5,  see Fig. \ref{fig1D}a). The accuracy of  our analytical predictions  for the mean FPT traces back to the quantitative  prediction for  the trajectories in the future of the FPT $\mu(t)$,  as shown in Fig. \ref{fig1D},\ref{fig3}. The strong dependence of $\mu(t)$ on the starting point $x_0$, predicted by our approach and confirmed numerically, is a direct manifestation of the non-Markovian feature of the random walks.  
 Together, our results demonstrate and quantify the importance of memory effects in first-passage properties of non-Markovian random walks in confined geometry. 




\newpage

\noindent{\bf Supplementary Information} is linked to the online version of the paper at www.nature.com/nature.\\

\noindent{\bf Acknowledgements}.  This work was supported by  ERC grant FPTOpt-277998.\\

\noindent{\bf Author Contributions}. All authors contributed equally to this work. Corresponding authors : Olivier B\'enichou (benichou@lptmc.jussieu.fr) and Raphael Voituriez (voiturie@lptmc.jussieu.fr)\\

\noindent{\bf Author Information}.  Reprints and permissions information is available at www.nature.com/reprints. The authors declare no competing financial interests. Correspondence and requests for materials should be addressed to O.B. (benichou@lptmc.jussieu.fr) and R.V.  (voiturie@lptmc.jussieu.fr)\\

\section*{ }

\newpage

\section*{ Figure legends}

\subsection*{Figure 1}
{\bf Mean first-passage time of a random walker in confinement}. {\bf a}: What is the mean  time $\Tm$ needed for a random walker to reach a target in a confining volume  ? In this paper, we answer  this question for random walkers with memory. {\bf b}: In one dimension, the problem is to quantify the first-passage time to a target in the presence of a reflecting boundary. We show here that $\Tm$ is controlled by the average trajectory $\mu(\tau)$ followed by the walker  in the  future of its first passage to the target.

\subsection*{Figure 2}
{\bf MFPT of 1-dimensional non Markovian random walks.}    MFPT as a function of the initial position $x_0$ ({\bf a-d}) and   average reactive trajectory $\mu(t)$ in the future of the FPT as a function of time $t$ ({\bf e-h}) for various one-dimensional Gaussian stochastic processes. Solid lines: predictions of the non-Markovian theory from Eqs.~(\ref{key1},\ref{key2}); dashed lines: Markovian approximation (in which $\mu(t)=0$); symbols: numerical simulations using the circulant matrix algorithm. In {\bf a,e}: Correlator $\Psi(t)$ as indicated with $D=1,\ D_0=30,\ \lambda=1$ (arbitrary units). Time is in unit of $1/\lambda$ and lengths in unit of $(D/\lambda)^{1/2}$. In {\bf e} different symbols represent different volumes (hexagones $V=40$, squares $V=60$, diamonds $V=120$); the superposition confirms that $\mu(t)$ does not depend on $V$. In {\bf b-d,f-h}: FBM with $K=1$ (arbitrary units).  Time is in unit of $V^{1/H}/K^{1/2H}$. Note that 
the theory is derived in the limit of large volume, or equivalently $x_0\ll V$. When significant,  error bars give the s.e.m. of the numerical simulations. Number $n$ of simulated trajectories : in {\bf a,e} $n=173285$ ($V=40$), $n=180641$ ($V=60$), $n=96623$ ($V=120$), in {\bf b,f} $n=19224$, in {\bf c,g} $n=22422$, in {\bf d ,h} $n=40685$.

\subsection*{Figure 3}
{\bf MFPT of 2 and 3 dimensional non Markovian random walks.}   MFPT to a target of radius $a=1$ (arbitrary units) as a function of the initial position $r_0$ ({\bf a,c,e}) and  average reactive trajectory $\mu(t)$ in the future of the FPT as a function of time $t$ ({\bf b,d,f}) for different 2-dimensional ({\bf a,b,c,d}) and  3-dimensional ({\bf e,f}) Gaussian stochastic processes.  Solid lines: predictions of the non-Markovian theory from Eqs.~(\ref{key1},\ref{key2}); dashed lines: Markovian approximation, in which $\mu(t)$ remains equal to the radius $a=1$  of the target; symbols: numerical simulations using the circulant matrix algorithm.   In {\bf a,b}: Correlator $\Psi(t)$ in 2D as indicated with $D=1,\ D_0=30, \ V=100, \ \lambda=1$ (arbitrary units). Time is in unit of $1/\lambda$ and lengths in unit of $a$. In {\bf c,d}: FBM in 2D with $K=1,\ V=60^2$ (arbitrary units). Time is in unit of $ a^{1/H}/K^{1/2H}$ and lengths in unit of $a$.  In {\bf e,f}: Correlator $\Psi(t)$ in 3D as indicated with $D=1,\ D_0=10, \lambda=1$ (arbitrary units). Time is in unit of $1/\lambda$ and lengths in unit of $a$. The confining volume is a sphere of radius $R=70$ or a cube of volume $V=116^3$. When significant,  error bars give the s.e.m. of the numerical simulations.  Number $n$ of simulated trajectories : in {\bf a,b} $n=35334$, in {\bf c,d} $n=37314$, in {\bf e,f} $n=16900$.

\newpage

\vspace{10cm}

\newpage

\vspace{10cm}

 \newpage

\end{document}